\documentclass[conference]{IEEEtran}
\IEEEoverridecommandlockouts
\usepackage{cite}
\usepackage{amsmath,amssymb,amsfonts}
\usepackage{algorithmic}
\usepackage{graphicx}
\usepackage{textcomp}
\usepackage{xcolor}
\def\BibTeX{{\rm B\kern-.05em{\sc i\kern-.025em b}\kern-.08em
    T\kern-.1667em\lower.7ex\hbox{E}\kern-.125emX}}

\usepackage[linesnumbered,ruled,vlined]{algorithm2e}
\usepackage{listings}
\usepackage[hidelinks]{hyperref}

\usepackage{tikz}
\usepackage{amsmath}
\usepackage{filecontents}

\usepackage{comment}
\usepackage[shortlabels]{enumitem}

\usepackage[linesnumbered,ruled,vlined]{algorithm2e}
\usepackage{tabularx}
\usepackage{booktabs}

\usepackage{algorithmic}
\usepackage{graphicx}
\usepackage{textcomp}
\usepackage{xcolor}
\usepackage{amsfonts}

\usepackage{multirow}
\usepackage[normalem]{ulem}
\useunder{\uline}{\ul}{}

\usepackage{tikz}
\usepackage{amssymb}

\usepackage{graphicx}
\usepackage{float}
\usepackage{subfig}

\usepackage{url}
\usepackage{multirow}

\usepackage{hyphenat}
\usepackage{comment}

\newcommand{\colorcyan}[1]{{\textcolor{purple}{#1}}}

\newcommand{\system}{\texttt{DB4NFV}\xspace}

\newcommand{\apiInput}{\hyperref[symbol:apiInput]{\textsl{assignInputSource}}\xspace}
\newcommand{\apiOutput}{\hyperref[symbol:apiOutput]{\textsl{assignOutputTarget}}\xspace}
\newcommand{\apiState}{\hyperref[symbol:apiState]{\textsl{registerState}}\xspace}
\newcommand{\apiStateObj}{\hyperref[symbol:apiStateObj]{\textsl{addStateObject}}\xspace}
\newcommand{\apiStateAccess}{\hyperref[symbol:apiStateAccess]{\textsl{addStateAccess}}\xspace}
\newcommand{\apiTxn}{\hyperref[symbol:apiTxn]{\textsl{addTransaction}}\xspace}
\newcommand{\apiVNF}{\hyperref[symbol:apiVNF]{\textsl{addVNF}}\xspace}
\newcommand{\apiNormalUDF}{\hyperref[symbol:apiNormalUDF]{\textsl{setPerFlowUDF}}\xspace}
\newcommand{\apiTxnUDF}{\hyperref[symbol:apiTxnUDF]{\textsl{setCrossFlowUDF}}\xspace}
\newcommand{\apiGetState}{\hyperref[symbol:apiGetState]{\textsl{getStateField}}\xspace}
\newcommand{\apiSetState}{\hyperref[symbol:apiSetState]{\textsl{setStateField}}\xspace}
\newcommand{\apiAbort}{\hyperref[symbol:apiAbort]{\textsl{abortTxn}}\xspace}
\newcommand{\apiAddTopoNode}{\hyperref[symbol:apiAddTopoNode]{\textsl{addTopoNode}}\xspace}

\usepackage{tikz}

\usepackage{makecell}

\newcommand{\compact}{\vspace{-0pt}}

\usepackage{float}
\begin{document}



\title{A Database System for State Management in Stateful Network Service Function Chains [Vision]}

\author{\IEEEauthorblockN{Zhonghao Yang}
\IEEEauthorblockA{\textit{Information Systems Technology and Design} \\
\textit{Singapore University of Technology and Design}\\
Singapore \\
zhonghao\_yang@mymail.sutd.edu.sg}
\and
\IEEEauthorblockN{Shuhao Zhang}
\IEEEauthorblockA{\textit{School of Computer Science and Engineering} \\
\textit{Nanyang Technological University}\\
Singapore \\
shuhao.zhang@ntu.edu.sg}
}

\maketitle

\begin{abstract}
Network Function Virtualization (NFV) heralds a transformative era in network function deployment, enabling the orchestration of Service Function Chains (SFCs) for delivering complex and dynamic network services. Yet, the development and sustenance of stateful SFCs remain challenging, with intricate demands for usability in SFC development, performance, and execution correctness.
In this paper, we present \system, a database system designed to address these challenges. Central to \system is the integration of transactional semantics into the entire lifecycle of stateful SFC, a core idea that enhances all aspects of the system. This integration provides an intuitive and well-structured API, which greatly simplifies the development of stateful SFCs. Concurrently, transactional semantics facilitate the optimization of runtime performance by efficiently leveraging modern multicore architectures. Moreover, by encapsulating state operations as transactions, \system achieves robustness, even at the entire chain level, ensuring reliable operation across varying network conditions. 
Consequently, \system marks a substantial forward leap in NFV state management, leveraging transactional semantics to achieve a harmonious blend of usability, efficiency, and robustness, thus facilitating the effective deployment of stateful SFCs in contemporary network infrastructures.

\end{abstract}
\compact
\section{Introduction}
Network Function Virtualization (NFV) has brought about a paradigm shift in network architectures by transitioning from traditional, hardware-dependent networking functions to agile, software-driven Virtualized Network Functions (VNFs)~\cite{mijumbi2015network}. Central to this shift are stateful Service Function Chains (SFCs), where the management of dynamic states across interconnected VNFs becomes a pivotal concern. The effective handling of these states is crucial as it dictates the performance, correctness, reliability, and scalability of the SFCs in response to network dynamics~\cite{bremler2016openbox, meng2019micronf, palkar2015e2, rajagopalan2013split, gember2014opennf, woo2018elastic}. The complexity inherent in managing stateful SFCs is further amplified by the need to meet stringent correctness demands, such as Chain-Output Equivalence (COE)~\cite{khalid2019correctness}, highlighting the intricate and multifaceted nature of state management in the evolving landscape of NFV.

\begin{figure}[t]
    \centering
    \includegraphics[width=0.48\textwidth]{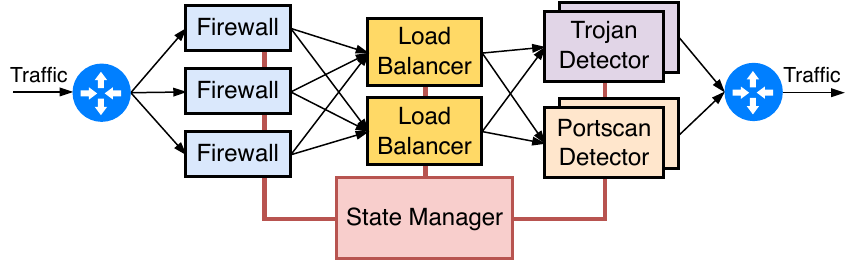}
    \caption{An Example of Stateful Service Function Chain with a Conceptual State Manager.}
    \label{fig:motivating_SFC}
\end{figure}

Figure~\ref{fig:motivating_SFC} depicts a representative stateful SFC composed of four types of network functions: a \textit{stateful firewall}, a \textit{load balancer}, a \textit{trojan detector}~\cite{de2014beyond}, and a \textit{portscan detector}~\cite{schechter2004fast}. Each VNF operates across multiple instances for enhanced performance and reliability. In this configuration, the firewall instances manage per-flow security states, essential for identifying and mitigating malicious activities, and operate primarily on local instance memory for per-flow state access. In contrast, the load balancer, trojan detector, and portscan detector necessitate collaborative management of shared states across multiple flows. For example, a load balancer instance, upon processing a new connection request, consults and updates the shared state to route the request to the optimally loaded host. 

The challenge of state management intensifies in scenarios involving network dynamics, such as scaling, load balancing, and fault tolerance, where states and traffic flows necessitate efficient redistribution or recovery among instances~\cite{khalid2019correctness}. The integration of a conceptual \textit{state manager} in this architecture abstracts data storage and concurrency control through well-defined interfaces. This abstraction enables instances to seamlessly access and manipulate state objects as needed, thereby obviating the requirement for frequent cross-instance state transfers. Such an approach not only isolates state management from VNF execution logic but also empowers developers to more effectively implement and adapt concurrency control and failure recovery strategies in response to evolving network conditions.

%
The concept of decoupling state management from NFV, despite being proposed in prior research~\cite{Kaplan2017StatelessNF}, presents three fundamental challenges that are yet to be comprehensively addressed. 
Firstly, the integration of increasingly complex network functions into modern SFCs necessitates NFV platforms that offer enhanced usability to streamline the development process. 
Secondly, the latency sensitivity inherent in network functions calls for advanced optimization techniques to ensure optimal performance of SFCs.
Thirdly, the diverse scopes and consistency requirements of network states within SFCs demand robust and unified management strategies, adaptable to the dynamic nature of network environments. 
While several frameworks~\cite{gember2014opennf, khalid2019correctness, LibVNF, pozza2021flexstate, woo2018elastic, rajagopalan2013pico, rajagopalan2013split} have endeavored to tackle these challenges, a \textbf{uniform solution} that simultaneously satisfies all three criteria remains elusive due to the intricate complexities involved in state management for SFCs. This gap highlights the pressing need for a state management solution that is not only flexible and scalable but also reliable, aligning with the continuous evolution and dynamic requirements of modern network infrastructures.

We introduce \system, a database system designed specifically for the nuanced requirements of state management in SFCs. Embracing the concept of decoupling state management from SFCs, \system incorporates transactional semantics into VNF state management. This integration is manifested through a suite of clear and intuitive transactional APIs, which markedly streamline the development process of stateful SFCs. 
\system leverages the capabilities of modern parallel processing architectures. It dynamically adjusts workload distribution and resource allocation strategies, thereby optimizing runtime performance of stateful SFCs in response to diverse traffic conditions. Additionally, \system ensures execution correctness across various scales, from individual VNFs to the entire SFC, thereby offering robust solutions for state consistency and fault tolerance, particularly in the face of network dynamism. 

The contributions of this paper are manifold and interlinked with the structure of the subsequent sections:

\begin{itemize}
    \item We present a thorough review of the state management landscape in NFV, introducing three key challenges that are fundamental to an effective state management system, and how existing works fail to address these challenges (Section~\ref{sec:motivation}). 
    \item To address the challenges, we propose \system, a database system designed specifically for NFV. The architectural nuances and key features of \system are elaborated in Section~\ref{sec:design}.
    \item The implementation details of \system, showcasing how its unified API is supported and integrated with existing NFV frameworks, together with some optimization details, is delineated in Section~\ref{sec:impl}. 
    \item We provide a conceptual analysis that summarizes and compares the features of existing works with those of \system in Section~\ref{sec:evaluation}.
    \item The paper concludes with a summary of our key contributions and a discussion of future work in Section~\ref{sec:conclusion}.
\end{itemize}

\compact
\section{Background and Motivation}
\label{sec:motivation}

\begin{figure*}
	\centering
	\begin{minipage}{\textwidth}
 
	\subfloat[VNF State Sharing]{
	       	\includegraphics[width=0.3\textwidth]{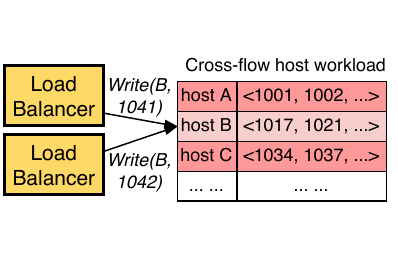}
            \label{fig:state_sharing}
        }	
        \subfloat[VNF Failure Isolation and Recovery]{   
    		\includegraphics[width=0.39\textwidth]{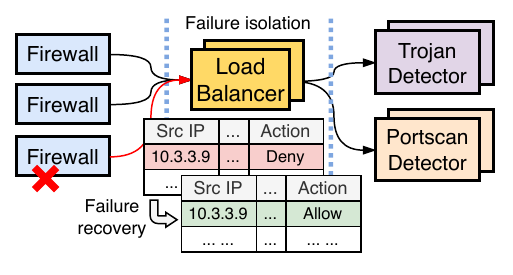}   
    		\label{fig:failure_recovery}
	}
	\subfloat[State Transfer under Network Dynamics]{
    		\includegraphics[width=0.3\textwidth]{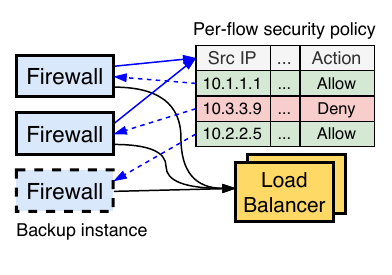}
            \label{fig:scaling}
	}	
	\end{minipage}
	\caption{Robust Stateful SFC Execution Illustration}
    \label{figures:correctness}
\end{figure*}

\subsection{SFC State Management Challenges}
\label{sec:features}
In the realm of stateful Service Function Chain (SFC) development, we delineate three pivotal challenges: \textit{usability}, \textit{efficiency}, and \textit{robustness}, each crucial to our work's foundation.

\noindent\textbf{Challenge 1: Enhancing Usability in Stateful SFC Development.} 
Usability is a key determinant in the development of stateful SFCs, where the complexity lies in encoding intricate VNF behaviors, ensuring accurate packet processing, and achieving high-performance execution under dynamic network conditions. Intuitive APIs are essential in this context, as they facilitate efficient VNF creation, management, and orchestration, enabling streamlined scaling and rapid integration of new functionalities into existing chains.

A significant aspect of usability concerns managing diverse state access scopes and execution logics across VNFs. Per-flow states, specific to individual traffic flows, are managed separately by respective instances, while cross-flow states involve shared data accessed and updated concurrently by multiple instances. Managing these concurrent state accesses, ensuring consistency and correct sequencing, is crucial for system integrity. The platform must adeptly handle the varying demands of per-flow and cross-flow states, catering to the wide array of network functions dependent on these state types.

\noindent\textbf{Challenge 2: Providing Efficient SFC Execution Runtime.} 
In stateful SFCs, ensuring data consistency and availability, particularly under variable network conditions, is crucial for effective state management~\cite{Kaplan2017StatelessNF,khalid2019correctness}. A common challenge arises during frequent read and update operations on cross-flow states, leading to synchronization conflicts. This issue is evident when multiple instances simultaneously modify a shared state, resulting in action blocks and downstream processing delays (as shown in Figure~\ref{fig:state_sharing}). For example, concurrent assignments by two load balancer instances to a single server without adequate synchronization lead to high contention and packet processing delays. Implementing an efficient state-sharing mechanism is therefore essential for optimizing stateful SFC performance.

Additionally, harnessing the capabilities of parallel architectures through strategic optimization techniques is fundamental to boost the overall performance and scalability of SFCs. Techniques such as state caching and workload balancing have been proposed to optimize VNF performance by reducing redundant state accesses and balancing loads~\cite{khalid2019correctness,woo2018elastic,pozza2021flexstate}. Nonetheless, advancing the execution of stateful SFCs requires addressing challenges that include adaptively responding to variable traffic workloads and maximizing multicore architecture utilization, all while considering the integrated structure of the SFC.

\noindent\textbf{Challenge 3: Ensuring Robustness in Stateful SFC.}
Robustness in stateful SFCs is imperative, particularly in handling VNF failures and network dynamics. Upon a VNF failure, the system must swiftly detect and nullify all state updates from the failed instance, as shown in Figure~\ref{fig:failure_recovery}. This action restores the system to a stable state and initiates failover procedures to maintain continuous traffic processing. Crucially, the recovery process must isolate the affected VNF to prevent disruptions in packet processing, state management, or routing decisions in interconnected VNFs.

Network traffic variability further complicates robustness in SFCs. Variations in traffic can necessitate scaling, load balancing, or straggler mitigation measures, involving complex redistributions of traffic and states across instances~\cite{khalid2019correctness,woo2018elastic,carvalho2022dyssect}. For instance, an increase in network traffic might require scaling up a stateful firewall (Figure~\ref{fig:scaling}), leading to redistribution of security policies and reallocation of network traffic for load balancing. Ensuring COE in such scenarios mandates careful management of state transfers and traffic allocations, ensuring no disruption to the normal functioning of other VNFs within the SFC while suppressing redundant operations.


\subsection{Related Works on VNF State Management}
The evolution of SFC deployment has been marked by the introduction of various VNF frameworks~\cite{gember2014opennf, Kaplan2017StatelessNF, meng2019micronf, LibVNF, khalid2019correctness, woo2018elastic}. Despite these developments, concurrently addressing \textit{usability}, \textit{efficiency}, and \textit{robustness} in state management continues to pose significant challenges.

\textbf{Lack of Uniform API.}
Current NFV frameworks, such as LibVNF~\cite{LibVNF}, OpenNF~\cite{gember2014opennf}, and S6~\cite{woo2018elastic}, offer APIs for constructing stateful VNFs. LibVNF provides essential interfaces for packet handling and data transmission, enabling fine-grained state exchange control across instances through event buffers and callbacks. OpenNF and S6 elevate state access abstraction from VNF execution, offering APIs that allow instances to retrieve and update network states with strong consistency. However, these frameworks fall short in supporting atomic updates, where multiple state changes must be executed or rolled back collectively. Moreover, their APIs are confined to individual NFs and do not encompass the coordination needed among multiple VNFs within an SFC. MicroNF~\cite{meng2019micronf} enables the consolidation and placement of modularized components across VNFs in an SFC, yet it does not support efficient management of per-flow and cross-flow states.
These limitations impose substantial coding complexities on developers, particularly in maintaining state consistency and ensuring execution correctness across stateful SFCs.

\textbf{Inefficient Execution Runtime.} 
While current NFV frameworks implement various concurrency control mechanisms to manage stateful operations, they often incur significant synchronization overhead, especially during high volumes of concurrent updates. Moreover, these frameworks generally do not fully leverage the benefits of parallel processing architectures in dynamic network environments. For instance, FlexState~\cite{pozza2021flexstate} assumes that shared states can be partitioned without synchronization, an approach that is not always feasible for VNFs with inherent state-sharing requirements.
OpenNF~\cite{gember2014opennf} introduces a two-phase state-sharing protocol, opting for state transfer across instances for eventual consistency or broadcasting updates for strong consistency as needed. S6~\cite{woo2018elastic} employs a distributed shared object model, periodically consolidating local updates into a global state. Similarly, CHC~\cite{khalid2019correctness} analyzes traffic workloads to determine the most suitable state-sharing technique, be it partitioning, caching, or operation offloading.
Despite these advancements in tailoring state-sharing strategies to network conditions, the existing solutions often rely on coarse-grained concurrency controls. This reliance results in significant locking overhead, particularly during frequent updates to shared states, thus adversely affecting the performance and scalability of stateful SFCs.

\textbf{Insufficient Support for Reliability.} Maintaining consistent shared states during concurrent accesses, coupled with ensuring accurate execution amidst network dynamics, presents a complex challenge. Most existing frameworks are tailored to support single NFs with specific network consistency requirements, but they fall short of addressing the broader spectrum of reliability needs in stateful SFCs. For instance, FTMB~\cite{sherry2015rollback} and Pico Replication~\cite{rajagopalan2013pico} provide failure recovery mechanisms for VNFs managing per-flow states. However, they do not adequately address the needs of VNFs dealing with shared cross-flow states. Frameworks such as OpenNF~\cite{gember2014opennf}, S6~\cite{woo2018elastic}, and FlexState~\cite{pozza2021flexstate} facilitate concurrent updates to shared states, but they lack mechanisms to ensure isolated execution and failure recovery for multiple VNFs within an SFC. 
Although CHC guarantees execution robustness, its performance suffers from a lock-based state-sharing mechanism under highly intensive concurrent state accesses.
As a result, developers utilizing these frameworks for stateful SFCs often face additional burdens to ensure COE and reliability, highlighting the need for more comprehensive solutions in this domain.

\subsection{The Need for a Unified Solution}
The stateful SFC domain grapples with significant state management challenges, primarily due to the varied and complex requirements of network functions. A critical issue stems from the tight coupling between state management APIs and their underlying data stores, which hampers adaptability and compounds development difficulties. Moreover, the absence of a standardized approach leads to inefficiencies, as developers are required to navigate multiple state management systems for different NFV scenarios. In light of these challenges, the demand for a unified, high-performance state management system for stateful SFCs is evident. Such a system, equipped with a singular API for various state access operations, would not only streamline development processes but also enhance overall system performance through efficient resource utilization and dynamic scheduling. A unified solution would ease the integration of VNFs across diverse operational contexts and simplify adherence to diverse reliability and consistency standards. 



\compact
\section{Key Designs}
\label{sec:design}
In this section, we begin by presenting the design philosophy of \system, offering an abstract perspective, followed by a detailed exploration of our system mechanisms that overcome the challenges of enhancing usability, execution efficiency, and offering a reliable execution environment.

%

\begin{figure}[t]
    \centering
    \includegraphics[width=0.48\textwidth]{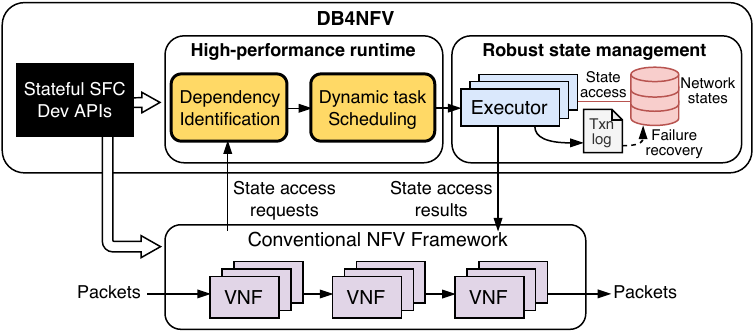} 
    \caption{Architectural Overview of the Conceptual Framework}
    \label{fig:sys_diagram}
\end{figure}

\subsection{The DB4NFV Abstraction}

\subsubsection{Design Philosophy}
A key idea of \system is expressing state access operations during SFC execution as database transactions. Encapsulating atomic stateful operations into a single transaction simplifies the declaration of complex VNF dependencies. By centralizing the control of transactional state accesses, concurrent access to shared states can be efficiently executed to enhance the overall performance. Furthermore, leveraging transaction ACID properties guarantees the reliability of the chain under network dynamics or execution failures.

\subsubsection{Architectural Overview}
Figure~\ref{fig:sys_diagram} provides an architectural overview of \system in conjunction with a standard NFV framework. \system introduces a set of user-friendly state access interfaces designed for seamless integration with existing NFV frameworks. Once initialized based on user specifications, the SFC is prepared to handle network traffic. At its core, \system features a centralized state manager responsible for overseeing network states and regulating state access requests from VNF instances, which are deployed on existing NFV frameworks.

\subsubsection{Execution Flow}
As illustrated in Figure~\ref{fig:sys_diagram}, the execution workflow of \system operates with NFV frameworks, where VNF instances process incoming packets according to per-flow logic. This entails reading packet headers, retrieving per-flow states from local memory, and determining packet forwarding paths. When VNF instances require access to cross-flow states, they forward these requests to the centralized state manager of \system. This manager organizes the requests into transactions, aligned with predefined user logic. 

\system's design ensures careful resolution of transactional dependencies prior to execution, scheduling these transactions across multiple executors for parallel processing. Each executor processes its set of transactions sequentially, effectively eliminating synchronization conflicts. Upon completion, \system relays the state access outcomes back to the VNF instances for further action, if necessary. The instances then proceed to forward the processed packets downstream, adhering to the established network topology.

\begin{table}[t]
\caption{DB4NFV API}
\label{tab:api}
\def\arraystretch{1.2}
\resizebox{0.48\textwidth}{!}{%
\begin{tabular}{|p{2.5cm}|p{6cm}|}
\hline
\textbf{Category} & \textbf{Function Name}\\ \hline

Network Configuration
& \apiInput(IP, port, protocol) \\ \cline{2-2}  
& \apiOutput(IP, port, protocol) \\ \cline{2-2}  
& \apiState(stateID, key, fields, access scope, consistency requirements) \\ \hline 
VNF state access templates 
& \apiStateObj(stateID, type) \\ \cline{2-2}
& \apiStateAccess(list of stateIDs, type) \\ \cline{2-2}
& \apiTxn(list of stateAccessIDs) \\ \hline
VNF execution logic 
& \apiVNF(list of txnIDs, normalUDFID, txnUDFID) \\ \cline{2-2}
& \apiNormalUDF(dataHolder) \\ \cline{2-2}
& \apiTxnUDF(dataHolder) \\ \hline
VNF State access operations 
& \apiGetState(stateID, field) \\ \cline{2-2}
& \apiSetState(stateID, field, value) \\ \cline{2-2}
& \apiAbort() \\ \hline
Network topology 
& \apiAddTopoNode(vnfID, parentID, stage, parallelism) \\ \hline
\end{tabular}%
}
\end{table}

\begin{algorithm}[t]
\footnotesize
  Job job = new Job(``newSFC"); \colorcyan{\tcp{\footnotesize{declare a new SFC}}}

   \colorcyan{\tcp{\footnotesize{declare StateObjects}}}

   \textbf{job.addStateObject(}``security\_policy"); 
 
  job.addStateObject(``host\_load"); 
  
  job.addStateObject(``request\_history"); 

  job.addStateObject(``portscan\_likelihood"); 

  \colorcyan{\tcp{\footnotesize{declare StateAccesses}}}

  \textbf{job.addStateAccess}(``update\_policy", ``security\_policy", ``W");
  
  job.addStateAccess(``update\_least\_loaded\_host", ``host\_load", ``W");
  
  job.addStateAccess(``evaluate\_traffic", ``request\_history", ``R");
  
  job.addStateAccess(``record\_activity", ``request\_history", ``W");

  job.addStateAccess(``check\_likelihood", ``portscan\_likelihood", ``R");

  job.addStateAccess(``update\_likelihood", ``portscan\_likelihood", ``W");

  \colorcyan{\tcp{\footnotesize{declare transactions}}}
  
  job.addTransaction(``lb\_txn", \{``update\_least\_loaded\_host"\});

  \textbf{job.addTransaction}(``td\_txn", \{``evaluate\_traffic", ``record\_activity"\});

  job.addTransaction(``ps\_txn", \{``check\_likelihood", ``update\_likelihood"\});

  \colorcyan{\tcp{\footnotesize{declare VNFs}}}
  
  job.addVNF(``firewall", fw\_perFlowUDF);

  job.addVNF(``load balancer", \{``lb\_txn"\}, lb\_perFlowUDF, lb\_crossFlowUDF);

  \textbf{job.addVNF}(``trojan detector", \{``td\_txn"\}, td\_perFlowUDF, td\_crossFlowUDF);

  job.addVNF(``portscan detector", \{``ps\_txn"\}, ps\_perFlowUDF, ps\_crossFlowUDF);

  \colorcyan{\tcp{\footnotesize{declare SFC topology}}}

  \textbf{job.addTopoNode}(``firewall", null, stage=1, parallelism=8);

  job.addTopoNode(``load balancer", ``firewall", stage=2, parallelism=8);

  job.addTopoNode(``trojan detector", ``load balancer", stage=3, parallelism=4);

  job.addTopoNode(``portscan detector", ``load balancer", stage=3, parallelism=4);
  
  job.start(); \colorcyan{\tcp{\footnotesize{Initialize SFC}}}
  \caption{Defining the example stateful SFC}
  \label{alg:api_example_SFC}
\end{algorithm}

\begin{algorithm}[t]
\footnotesize
\textbf{Function} IDS\_Cross\_Flow\_UDF (dataHolder):

requestHistory = \textbf{dataHolder.getStateField}(“request\_history”);

newRequest = dataHolder.getPacketData(“new\_request”);

isMalicious = securityCheck(requestHistory, newRequest);

\eIf(\colorcyan{\tcp*[h]{Raise alarm and abort txn}}){\text{isMalicious}}{
    
    notifyHost();
    
    \textbf{dataHolder.abortTxn}();
    
}(\colorcyan{\tcp*[h]{Record new request}}){

    \textbf{dataHolder.setStateField}(“request\_history”, requestHistory, newRequest); 
}

  \caption{API demonstration for defining cross-flow UDF for Trojan Detector}
  \label{alg:api_example_IDS}
\end{algorithm}

\subsection{Uniform API for Stateful SFC}
\label{sec:api_explanation}
To facilitate the SFC development, \system incorporates transactional semantics into stateful SFC declarations, providing users with a well-structured approach to defining complex VNF behaviors. The APIs are summarized in Table~\ref{tab:api}. 

\subsubsection{Network Configuration}
Functions~\apiInput and \apiOutput specify the source of input packets and targets to receive processed packets for stateful SFCs. Meanwhile, Function~\apiState configures the schema of network states, as well as their access scopes and consistency requirements. The declared states can be further referenced during VNF logic declarations.

\subsubsection{State Access Operation}
\label{sec:stateAccessOperationAPI}
Function~\apiStateObj and~\apiStateAccess allows users to define the state access operations by referencing configured network states. These state access operations can be further encapsulated into transactions via Function~\apiTxn, ensuring their atomicity based on VNF requirements.

\subsubsection{VNF Execution Logic}
\label{sec:api_vnf_execution}
Function~\apiVNF defines VNF execution logics, including their corresponding set of transactions, and two types of user-defined functions. Function~\apiNormalUDF specifies per-flow processing procedures that only access instance local memories, and Function~\apiTxnUDF specifies the cross-flow processing procedures to shared network states.

\subsubsection{Network Topology}
\system abstracts the topology of stateful SFCs as logical directed acyclic graphs (DAGs), whose nodes represent network functions and edges signify inter-VNF traffic flow. Function~\apiAddTopoNode allows users to declare the position of VNFs in the topology, as well as their level of parallelism during execution.

\textbf{Example of Developing SFC with \system.}
\label{sec:api_sfc_example}
To demonstrate the usability of our API, Algorithm~\ref{alg:api_example_SFC} shows the definition of the example stateful SFC. After registering a new SFC job, the user declares the state objects to be visited during state access operations. 
For example, the stateful firewall declares access to its security policy states using~\apiStateObj(``security\_policy"). 
State objects can be further combined to describe state access operations using Function~\apiStateAccess. \system supports two primary types of state access operations: (1) Read, representing a read action on a single state object, and (2) Write, representing a write action to a single state object, coupled with conditional read actions on multiple states. 

Atomic state access operations are encapsulated as a transaction via Function~\apiTxn. For instance, the transaction of the trojan detector contains two steps: (1) reading the host request history to evaluate a new request, and (2) updating the request history if no threats are detected. These two operations should be collectively executed.
The declared transactions are further assigned to their corresponding VNFs via Function~\apiVNF, along with two UDFs encoding VNF execution procedures. 
Finally, the stateful SFC is constructed by adding declared VNFs as topology nodes via Function~\apiAddTopoNode. The stateful firewall has no upstream node and its stage is set to 1, indicating its starting position in the chain. Meanwhile, both the trojan detector and the portscan detector specify load balancer as their common upstream node, and they have the same stage number of 3. Users can also configure the number of parallel executors to be deployed for each VNF.

To further illustrate, Algorithm~\ref{alg:api_example_IDS} shows how cross-flow UDFs can be defined for a trojan detector, which detects malicious patterns in packet request sequences to host resources (E.g., SSH connection, HTTP download, FTP download, IRC connection)~\cite{de2014beyond}. Upon receiving a new request, an instance first acquires the shared request history through~\apiGetState and conducts a security check. If the overall request sequence matches the malicious pattern, the instance will report to the host and abort this transaction using~\apiAbort. Otherwise, the new request is updated to the state by~\apiSetState.
Lastly, users can fine-tune the detailed execution settings of the \system system using a series of system configuration functions. These functions provide the flexibility to configure workload scheduling, logging, and benchmarking parameters, ensuring a tailored and optimized system configuration tailored to specific requirements.

\begin{figure}[t]
    \centering
    \includegraphics[width=0.42\textwidth]{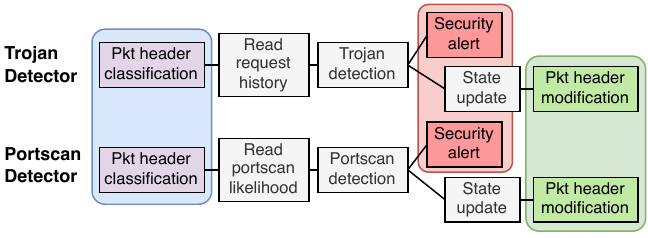}
    \caption{Modularization on two IDSs}
    \label{fig:modularization}
\end{figure}

\subsection{High-Performance Runtime}
\label{sec:runtime}
Transactional semantics integration into VNF shared state management enhances the development of stateful SFCs and improves shared state management with well-defined transactional dependencies. However, mere transactional representation of state access operations is insufficient to mitigate synchronization conflicts under concurrent accesses. \system, therefore, employs multiple optimization techniques to leverage parallel architecture effectively and boost SFC performance.

\textbf{Adaptive Transaction Workload Scheduling.} \system identifies the fine-grained workload dependencies among state access operations before execution. These processes are performed based on stage, indicating the position of VNFs in the SFC (Section~\ref{sec:api_sfc_example}). Stage restricts the boundary for parallel execution. Transaction requests from the same stage can be processed concurrently, or they must be sequentially executed. Upon receiving state access requests, \system constructs one Task Precedence Graph (TPG) for each stage, where nodes symbolize state access operations and edges represent transactional dependencies. Dependencies are categorized as time-based (for operations accessing the same state object at different times), parametric (where one operation depends on another's result), or logical (defining transaction boundaries). The TPG informs the allocation of workloads to parallel threads.

Once the TPG is constructed, \system allocates state access workloads to parallel threads. Given the highly dynamic and unpredictable nature of network traffic workload characteristics, \system adaptively selects the optimal task scheduling strategy from a pool of schedulers, aiming to fully optimize multicore resources and enhance overall system performance. The schedulers vary based on the graph exploration algorithms (BFS, DFS, Non-Structured), or whether to group multiple state access operations to trade off between improving scalability with fine-grained task allocation and reducing context-switching overhead. A heuristic decision model guides the selection of the optimal strategy based on multiple criteria, including the distribution of three types of dependencies, the skewness among state access operations, and an estimation of their computational complexities.

\textbf{Support of VNF Modularization.} \system also incorporates VNF modularization~\cite{meng2019micronf, bremler2016openbox, kohler2000click}, segmenting VNF logic into discrete modules. Figure~\ref{fig:modularization} illustrates applying modularization on the trojan detector and portscan detector, highlighting their common modular that can be reused to support both VNFs. During SFC declaration, \system allows users to define their VNFs in the form of collective VNF modular. Based on their functionalities and traffic dependencies, \system identifies and determines the feasibility of reusing modular components to support multiple VNFs in the chain, and generates an optimal modular placement strategy among instances. The placement strategy is determined so that different cores have minimum communications to support the collective execution of modular.

\textbf{Caching of Infrequently Updated States.} In scenarios where VNFs or specific network traffic patterns predominantly involve frequent Read operations on shared states with infrequent updates, \system implements a strategic caching mechanism. This approach is geared towards reducing redundant state accesses and minimizing packet processing delays. \system evaluates the ratio of Read requests against the total transaction batch. When the Read operations significantly outweigh updates, surpassing a predetermined threshold, \system enhances performance by caching these states in each thread's local memory. This cached information is promptly synchronized with the central state upon any update to maintain consistency. In contrast, states experiencing a balanced or high ratio of updates are classified as frequently updated and are retained in the centralized datastore. Here, they are managed via \system's transactional concurrency control mechanism, ensuring synchronized and efficient state access across the system.

\begin{table*}[]
\caption{Comparison with Existing NFV Frameworks}
\label{tab:compare}
\def\arraystretch{1.1}
\begin{tabular}{|c|ccc|cccc|cc|}
\hline
\multirow{2}{*}{\begin{tabular}[c]{@{}c@{}}Stateful NFV \\ Frameworks\end{tabular}} & \multicolumn{3}{c|}{\begin{tabular}[c]{@{}c@{}}Usability in Stateful \\ SFC Development\end{tabular}}                           & \multicolumn{4}{c|}{\begin{tabular}[c]{@{}c@{}}Efficient SFC \\ Execution Runtime\end{tabular}}                                   & \multicolumn{2}{c|}{\begin{tabular}[c]{@{}c@{}}Robustness in \\ Stateful SFC\end{tabular}}  \\ \cline{2-10} 
& \multicolumn{1}{c|}{\begin{tabular}[c]{@{}c@{}}Support for \\ Varying \\ State Scopes\end{tabular}} & \multicolumn{1}{c|}{\begin{tabular}[c]{@{}c@{}}Support for \\ State Access \\ Atomicity\end{tabular}} & \begin{tabular}[c]{@{}c@{}}Support for \\ VNF \\ Coordination\end{tabular} & \multicolumn{1}{c|}{\begin{tabular}[c]{@{}c@{}}Concurrent \\ State Access \\ Efficiency\end{tabular}} & \multicolumn{1}{c|}{\begin{tabular}[c]{@{}c@{}}Failure \\ Recovery \\ Efficiency\end{tabular}} & \multicolumn{1}{c|}{\begin{tabular}[c]{@{}c@{}}State \\ Migration \\ Efficiency\end{tabular}} & \begin{tabular}[c]{@{}c@{}}Utilization of \\ Multicore \\ Architectures\end{tabular} & \multicolumn{1}{c|}{\begin{tabular}[c]{@{}c@{}}Failure \\ Recovery \\ Reliability\end{tabular}} & \begin{tabular}[c]{@{}c@{}}State \\ Migration \\ Reliability\end{tabular} \\ \hline
FTMB~\cite{sherry2015rollback}                                                                               & \multicolumn{1}{c|}{$\times$}                                                                             & \multicolumn{1}{c|}{$\times$}                                                                                 & $\times$                                                                           & \multicolumn{1}{c|}{$\times$}                                                                                 & \multicolumn{1}{c|}{$\checkmark$$\checkmark$}                                                                          & \multicolumn{1}{c|}{$\times$}                                                                         & $\times$                                                                                    & \multicolumn{1}{c|}{$\times$}                                                                           & $\times$                                                                          \\ \hline
OpenNF~\cite{gember2014opennf}                                                                             & \multicolumn{1}{c|}{$\checkmark$}                                                                               & \multicolumn{1}{c|}{$\times$}                                                                                 & $\times$                                                                           & \multicolumn{1}{c|}{$\times$}                                                                                 & \multicolumn{1}{c|}{$\checkmark$}                                                                          & \multicolumn{1}{c|}{$\times$}                                                                         & $\times$                                                                                     & \multicolumn{1}{c|}{$\times$}                                                                           & $\checkmark$                                                                          \\ \hline
StatelessNF~\cite{Kaplan2017StatelessNF}                                                                             & \multicolumn{1}{c|}{$\checkmark$}                                                                               & \multicolumn{1}{c|}{$\times$}                                                                                 & $\times$                                                                           & \multicolumn{1}{c|}{$\checkmark$}                                                                                 & \multicolumn{1}{c|}{$\checkmark$}                                                                          & \multicolumn{1}{c|}{$\checkmark$}                                                                         & $\times$                                                                                     & \multicolumn{1}{c|}{$\times$}                                                                           & $\checkmark$                                                                          \\ \hline
S6~\cite{woo2018elastic}                                                                                 & \multicolumn{1}{c|}{$\checkmark$}                                                                               & \multicolumn{1}{c|}{$\times$}                                                                                 & $\times$                                                                           & \multicolumn{1}{c|}{$\times$}                                                                                 & \multicolumn{1}{c|}{$\times$}                                                                          & \multicolumn{1}{c|}{$\checkmark$}                                                                         & $\checkmark$                                                                                     & \multicolumn{1}{c|}{$\times$}                                                                           & $\checkmark$                                                                          \\ \hline
CHC~\cite{khalid2019correctness}                                                                                & \multicolumn{1}{c|}{$\checkmark$ }                                                                               & \multicolumn{1}{c|}{$\checkmark$}                                                                                 & $\checkmark$                                                                            & \multicolumn{1}{c|}{$\times$}                                                                                 & \multicolumn{1}{c|}{$\checkmark$}                                                                          & \multicolumn{1}{c|}{$\checkmark$}                                                                         & $\checkmark$                                                                                      & \multicolumn{1}{c|}{$\checkmark$}                                                                           & $\checkmark$                                                                           \\ \hline
MicroNF~\cite{meng2019micronf}                                                                                & \multicolumn{1}{c|}{$\times$}                                                                               & \multicolumn{1}{c|}{$\times$}                                                                                 & $\checkmark$                                                                            & \multicolumn{1}{c|}{$\times$}                                                                                 & \multicolumn{1}{c|}{$\times$}                                                                          & \multicolumn{1}{c|}{$\times$}                                                                         & $\checkmark$                                                                                      & \multicolumn{1}{c|}{$\times$}                                                                           & $\times$                                                                           \\ \hline
FlexState~\cite{pozza2021flexstate}                                                                                & \multicolumn{1}{c|}{$\checkmark$ }                                                                               & \multicolumn{1}{c|}{$\times$}                                                                                 & $\times$                                                                            & \multicolumn{1}{c|}{$\times$}                                                                                 & \multicolumn{1}{c|}{$\checkmark$}                                                                          & \multicolumn{1}{c|}{$\times$}                                                                         & $\checkmark$                                                                                      & \multicolumn{1}{c|}{$\times$}                                                                           & $\times$                                                                           \\ \hline
\textbf{DB4NFV}                                                                             & \multicolumn{1}{c|}{$\checkmark$}                                                                               & \multicolumn{1}{c|}{$\checkmark$$\checkmark$}                                                                                 & $\checkmark$$\checkmark$                                                                            & \multicolumn{1}{c|}{$\checkmark$$\checkmark$}                                                                                 & \multicolumn{1}{c|}{$\checkmark$}                                                                          & \multicolumn{1}{c|}{$\checkmark$}                                                                         & $\checkmark$$\checkmark$                                                                                     & \multicolumn{1}{c|}{$\checkmark$}                                                                           & $\checkmark$                                                                          \\ \hline
\end{tabular}
\end{table*}

\subsection{Robust State Management}
\label{sec:state_management}
Robustness in SFC execution is crucial, necessitating network states to be consistent despite execution failures or fluctuating network conditions. In \system, per-flow states, as well as states amenable to partitioning or infrequent shared access, are managed as local states within each instance's memory for rapid access. Conversely, cross-flow states frequently accessed by multiple instances are centralized in a key-value database, with a state manager overseeing access.

\textbf{Fault Tolerance Execution.} \system’s decoupled state management architecture enhances fault tolerance, effectively isolating failures within the job. To support this, \system utilizes multi-version state storage. State access executors update snapshots and transaction histories at regular intervals, laying the groundwork for quick state recovery post-failures. During operation, \system records interim results from dependency resolutions, including potential transaction aborts and parametric dependencies, where one state access hinges on another's update. In the event of an instance failure, \system promptly identifies and halts all impacted state operations, guided by its dependency logs. It then reverts to the most recent stable state snapshot, ensuring the system's consistency is preserved.

\textbf{Robustness under Network Dynamics.} The decoupled state management approach equips \system with the agility to uphold state consistency amidst diverse network conditions, while also ensuring scalability. Under regular operation, per-flow states are maintained within local memory for efficient access and modification. However, in response to dynamic network conditions necessitating state migration, these local states are temporarily shifted to the centralized database, becoming accessible shared resources for other relevant instances. For example, if a VNF is required to scale up due to a surge in network traffic, the per-flow states from existing instances are transferred to the centralized database. The state manager then reallocates these states among newly provisioned instances. Similarly, when a straggler VNF instance is detected, its local states are registered in the database and shared with a backup instance to ensure continuity. Upon completion of such migrations, the temporarily shared states are removed from the centralized database and reverted to their original locations in local memories. This method of state management fortifies the robustness of \system in handling network dynamics, seamlessly adapting to changing conditions without compromising system stability.

\compact
\section{Implementation Details}
\label{sec:impl}
\begin{figure}[t]
    \centering
    \includegraphics[width=0.45\textwidth]{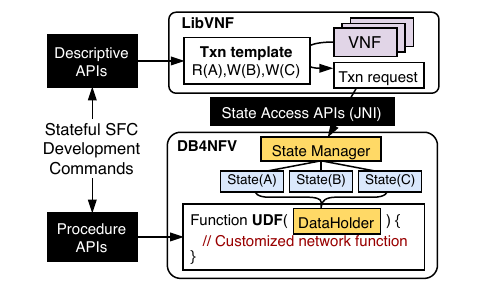}
    \caption{Implementation of DB4NFV}
    \label{fig:implementation}
\end{figure}

This section discusses more implementation details of \system, outlined in Figure~\ref{fig:implementation}.
It is publicly accessible\footnote{\url{https://github.com/intellistream/MorphStream/tree/DB4NFV}}. 

\textbf{API Implementation Details.} 
\system significantly simplifies the development of stateful SFCs through its structured client-side APIs, divided into descriptive and procedural categories. The \textit{descriptive APIs} provide users the tools to define the SFC's architecture, including state schemas, operational processes, and network topologies, utilizing functions like \apiStateAccess and \apiTxn. On the other hand, \textit{procedural APIs} concentrate on the executable aspects of VNFs, specifically targeting user-defined functions as elaborated in Section~\ref{sec:api_vnf_execution}.

\textbf{Integration with NFV Frameworks.} 
Built upon a recent transactional stream processing system~\cite{10.1145/3588913}, \system features transactional state access APIs that facilitate seamless integration with various NFV frameworks. A notable integration achievement is with libVNF~\cite{LibVNF}, which specializes in packet transmission and kernel-bypass optimizations. This integration uses the Java Native Interface for effective data transfer and API communication between Java and C++ environments. During runtime, VNF instances in libVNF process incoming packets and conduct per-flow operations. For cross-flow state access, these instances send transactional state access requests to \system. These requests, treated as callback functions, allow for the continuation of per-flow processing while \system handles the state access.

\textbf{Data Encoding and Template Utilization.} 
To optimize performance in managing complex transactional structures, \system adopts a byte stream encoding strategy for state access requests, focusing solely on packet-specific data to minimize processing overhead. Additionally, \system utilizes static descriptive templates to store common information, readily available to all executors. This approach of processing byte streams, where \system extracts essential packet data and transactional dependencies from these templates, leads to efficient execution of user-defined functions. Such a streamlined process eases development challenges, enabling developers to manage state access with clarity and efficiency in their data structures.

\compact
\section{Comparing to Existing Works}
\label{sec:evaluation}
In the development of \system, we encountered a unique challenge in conducting a direct empirical comparison with existing solutions. Current state management solutions for SFCs lack a unified approach that encapsulates all the features necessary for a comprehensive evaluation. This absence of a holistic solution in the market necessitates the re-implementation of these various solutions within the \system framework to enable a meaningful comparison. Given the extensive scope of such an endeavor, coupled with the visionary nature of this paper, we have focused on presenting a conceptual analysis rather than empirical results at this stage.

To provide a clear perspective on the current state of the field and the positioning of \system, we have prepared a detailed table as shown in Table~\ref{tab:compare} that summarizes and compares the features of existing works with those of \system. This comparative summary underscores the unique contributions of \system and highlights its potential to address the gaps and limitations present in current NFV technologies.

Although varying scopes of network states are supported by most existing works, they either ignore or fail to provide an intuitive interface for the declaration of state access atomicity and the coordination across VNFs in the chain. In contrast, \system provides well-structured APIs to efficiently support the development of stateful SFCs.
Moreover, despite existing efforts in optimizing stateful SFC execution performance, they all suffer from high synchronization overhead during concurrent state accesses and fail to optimize the usage of multicore architectures. \system eliminates contention overhead by fine-grained dependency resolution before state access execution, and adaptively adjusting its task scheduling strategies to parallel executors based on real-time traffic. It further enhances SFC scalability with various optimization techniques leveraging multicore architectures.
Lastly, there exist gaps in most existing NFV frameworks in ensuring the robustness of SFC, as they lack the support for enforcing state access atomicity under failures. In response, \system guarantees atomic state access execution with transactional semantics and provides a robust SFC execution environment prone to VNF failures and network dynamics.
To summarize, \system offers a uniform solution to address the challenges in stateful SFCs simultaneously.

\compact
\section{Conclusion}
\label{sec:conclusion}
In this paper, we introduced \system, a database system uniquely tailored for the complex requirements of managing state in stateful SFCs. By integrating transactional semantics into VNF state management, \system simplifies the development process and enhances the management of shared states. Its architecture adeptly addresses critical NFV challenges, including synchronization conflicts and efficient use of multicore architectures, thereby ensuring robust performance across dynamic network conditions and in scenarios of VNF failures. Through novel workload scheduling and fault tolerance approaches, \system markedly improves the scalability and reliability of NFV platforms. 
Looking to the future, \system lays a solid foundation for ongoing data-centric innovations in NFV technology. We plan to further refine state management techniques within \system and expand its adaptability to a wider range of network conditions and use cases. The promise of \system reaches beyond its current capabilities, positioning it as a catalyst for innovation and a significant contributor to the evolution of modern network infrastructures.

\compact

\bibliographystyle{ieeetr} 
\bibliography{reference}

\begin{thebibliography}{10}

\bibitem{mijumbi2015network}
R.~Mijumbi, J.~Serrat, J.-L. Gorricho, N.~Bouten, F.~De~Turck, and R.~Boutaba, ``Network function virtualization: State-of-the-art and research challenges,'' {\em IEEE Communications surveys \& tutorials}, vol.~18, no.~1, pp.~236--262, 2015.

\bibitem{bremler2016openbox}
A.~Bremler-Barr, Y.~Harchol, and D.~Hay, ``Openbox: A software-defined framework for developing, deploying, and managing network functions,'' in {\em Proceedings of the 2016 ACM SIGCOMM Conference}, pp.~511--524, 2016.

\bibitem{meng2019micronf}
Z.~Meng, J.~Bi, H.~Wang, C.~Sun, and H.~Hu, ``Micronf: An efficient framework for enabling modularized service chains in nfv,'' {\em IEEE Journal on Selected Areas in Communications}, vol.~37, no.~8, pp.~1851--1865, 2019.

\bibitem{palkar2015e2}
S.~Palkar, C.~Lan, S.~Han, K.~Jang, A.~Panda, S.~Ratnasamy, L.~Rizzo, and S.~Shenker, ``E2: A framework for nfv applications,'' in {\em Proceedings of the 25th Symposium on Operating Systems Principles}, pp.~121--136, 2015.

\bibitem{rajagopalan2013split}
S.~Rajagopalan, D.~Williams, H.~Jamjoom, and A.~Warfield, ``Split/merge: system support for elastic execution in virtual middleboxes.,'' in {\em NSDI}, vol.~13, pp.~227--240, 2013.

\bibitem{gember2014opennf}
A.~Gember-Jacobson, R.~Viswanathan, C.~Prakash, R.~Grandl, J.~Khalid, S.~Das, and A.~Akella, ``Opennf: Enabling innovation in network function control,'' {\em ACM SIGCOMM Computer Communication Review}, vol.~44, no.~4, pp.~163--174, 2014.

\bibitem{woo2018elastic}
S.~Woo, J.~Sherry, S.~Han, S.~Moon, S.~Ratnasamy, and S.~Shenker, ``Elastic scaling of stateful network functions,'' in {\em 15th $\{$USENIX$\}$ Symposium on Networked Systems Design and Implementation ($\{$NSDI$\}$ 18)}, pp.~299--312, 2018.

\bibitem{khalid2019correctness}
J.~Khalid and A.~Akella, ``Correctness and performance for stateful chained network functions.,'' in {\em NSDI}, vol.~19, pp.~26--28, 2019.

\bibitem{de2014beyond}
L.~De~Carli, R.~Sommer, and S.~Jha, ``Beyond pattern matching: A concurrency model for stateful deep packet inspection,'' in {\em Proceedings of the 2014 ACM SIGSAC Conference on Computer and Communications Security}, pp.~1378--1390, 2014.

\bibitem{schechter2004fast}
S.~E. Schechter, J.~Jung, and A.~W. Berger, ``Fast detection of scanning worm infections,'' in {\em Recent Advances in Intrusion Detection: 7th International Symposium, RAID 2004, Sophia Antipolis, France, September 15-17, 2004. Proceedings 7}, pp.~59--81, Springer, 2004.

\bibitem{Kaplan2017StatelessNF}
M.~Kaplan, A.~Alsudais, E.~Keller, and F.~Le, ``Stateless network functions: Breaking the tight coupling of state and processing,'' in {\em Symposium on Networked Systems Design and Implementation}, 2017.

\bibitem{LibVNF}
P.~Naik, A.~Kanase, T.~Patel, and M.~Vutukuru, ``Libvnf: Building virtual network functions made easy,'' in {\em Proceedings of the ACM Symposium on Cloud Computing}, SoCC '18, (New York, NY, USA), p.~212–224, Association for Computing Machinery, 2018.

\bibitem{pozza2021flexstate}
M.~Pozza, A.~Rao, D.~F. Lugones, and S.~Tarkoma, ``Flexstate: Flexible state management of network functions,'' {\em IEEE Access}, vol.~9, pp.~46837--46850, 2021.

\bibitem{rajagopalan2013pico}
S.~Rajagopalan, D.~Williams, and H.~Jamjoom, ``Pico replication: A high availability framework for middleboxes,'' in {\em Proceedings of the 4th annual Symposium on Cloud Computing}, pp.~1--15, 2013.

\bibitem{carvalho2022dyssect}
F.~B. Carvalho, R.~A. Ferreira, {\'I}.~Cunha, M.~A. Vieira, and M.~K. Ramanathan, ``Dyssect: Dynamic scaling of stateful network functions,'' in {\em IEEE INFOCOM 2022-IEEE Conference on Computer Communications}, pp.~1529--1538, IEEE, 2022.

\bibitem{sherry2015rollback}
J.~Sherry, P.~X. Gao, S.~Basu, A.~Panda, A.~Krishnamurthy, C.~Maciocco, M.~Manesh, J.~Martins, S.~Ratnasamy, L.~Rizzo, {\em et~al.}, ``Rollback-recovery for middleboxes,'' in {\em Proceedings of the 2015 ACM Conference on Special Interest Group on Data Communication}, pp.~227--240, 2015.

\bibitem{kohler2000click}
E.~Kohler, R.~Morris, B.~Chen, J.~Jannotti, and M.~F. Kaashoek, ``The click modular router,'' {\em ACM Transactions on Computer Systems (TOCS)}, vol.~18, no.~3, pp.~263--297, 2000.

\bibitem{10.1145/3588913}
Y.~Mao, J.~Zhao, S.~Zhang, H.~Liu, and V.~Markl, ``Morphstream: Adaptive scheduling for scalable transactional stream processing on multicores,'' {\em Proc. ACM Manag. Data}, vol.~1, may 2023.

\end{thebibliography}

\end{document}